\documentclass{ws-ijmpc}
\newcommand{\ti}{\textit}
\newcommand{\be}{\begin{equation}}
\newcommand{\ee}{\end{equation}}
\newcommand{\bea}{\begin{eqnarray}}
\newcommand{\eea}{\end{eqnarray}}
\newcommand{\la}{\langle}
\newcommand{\ra}{\rangle}

\begin{document}

\markboth{Andrea Baronchelli, Vittorio Loreto and Luc Steels}
{In-depth analysis of the Naming Game dynamics: the homogeneous mixing case}

\catchline{}{}{}{}{}

\title{In-depth analysis of the Naming Game dynamics: \\ the homogeneous mixing case
}

\author{ANDREA BARONCHELLI\footnote{Corresponding author.}}

\address{Departament de F\'isica i Enginyeria Nuclear, \\
Universitat Polit\`ecnica de Catalunya, \\
Campus Nord, M\'odul B4, \\
08034 Barcelona, Spain \\
andrea.baronchelli@upc.edu}

\author{VITTORIO LORETO}

\address{Dipartimento di Fisica \\
 Universit\`a ``La Sapienza'' and SMC-INFM \\ 
 P.le A. Moro 2, 00185 Roma, Italy \\
and Complex Networks Lagrange Laboratory, ISI Foundation, \\
Viale S. Severo 65, 10133, Torino, Italy \\ 
 vittorio.loreto@roma1.infn.it
}

\author{LUC STEELS}

\address{VUB AI Lab, Brussels, Belgium \\
 Sony Computer Science Laboratory, Paris, France \\
 steels@arti.vub.ac.be 
}

\maketitle
\begin{history}
\received{Day Month Year}
\revised{Day Month Year}
\end{history}

\begin{abstract}
  Language emergence and evolution has recently gained growing
  attention through multi-agent models and mathematical frameworks to
  study their behavior. Here we investigate further the Naming Game, a
  model able to account for the emergence of a shared vocabulary of
  form-meaning associations through social/cultural learning. Due to
  the simplicity of both the structure of the agents and their
  interaction rules, the dynamics of this model can be analyzed in
  great detail using numerical simulations and analytical
  arguments. This paper first reviews some existing results and then
  presents a new overall understanding.
\end{abstract}

\section{Introduction}

Language is based on a set of cultural conventions socially shared by
a group. But how are these conventions established without a central
coordinator and without telepathy? The problem has been addressed by
several disciplines, but it is only in the last decade that there has
been a growing effort to tackle it scientifically using multi-agent
models and mathematical approaches
(cfr.\cite{NowakKrak1999,steels_connection,castellano07} for a
review). Initially these models focused on the emergence of a shared
vocabulary, but increasingly attempts are made to tackle
grammar~\cite{NowakKrak1999,nowak_synt,brighton01,puglisi07,steels98origins}.

The proposed models can be classified as defending a sociobiological
or a sociocultural explanation. The sociobiological
approach~\cite{hurford89}, which includes the evolutionary language
game~\cite{NowakKrak1999}, is based on the assumption that successful
communicators, enjoying a selective advantage, are more likely to
reproduce than worse communicators. If communication strategies are
innate, then more successful strategies will displace rivals.  The
term strategy acquires its precise meaning in the context of a
particular model. For instance, it can be a strategy for acquiring the
lexicon of a language, i.e., a function from samplings of observed
behaviors to acquired communicative behavior
patterns~\cite{hurford89,oliphant_batali,Nowak1999}, or it can simply
coincide with the lexicon of the parents~\cite{NowakKrak1999} or with
some strong disposition to acquire a particular kind of syntax,
usually called innate Universal Grammar~\cite{chomsky88}.

In this paper we discuss a model, first proposed
in~\cite{baronchelli_ng_first}, that belongs to the sociocultural
family~\cite{Hutchins95,steels1995,lenaerts2005}. Here, good
strategies do not necessarily provide higher reproductive success, but
only higher communicative success and greater expressive power, and
hence greater success in reaching cooperative goals, with less
effort. Agents select better strategies exploiting cultural choices,
feedback from communication, and a sense of effort. Agents have not
only the ability to acquire an existing system but to expand their
rules to deal with new communicative challenges and to adjust their
rules based on observing the behavior of others.  Global coordination
emerges over cultural timescales, and language is seen as an evolving
and self-organized system~\cite{steels_dynsys}. While the
sociobiological approach emphasizes language transmission following a
vertical, genetic or generational line, the sociocultural approach
emphasizes peer-to-peer interaction~\cite{cavalli-sforza81book}.

A second, fundamental distinction among the different models concerns
the adopted mechanisms of social learning describing how stable
dispositions are exchanged and coordinated between
individuals~\cite{boyd1985}. The two main approaches are the so called
observational learning model and the reinforcement
model~\cite{rosenthal1978}. In the first
approach~\cite{hurford89,oliphant_batali,Nowak1999,NowakKrak1999},
observation is the main ingredient of learning and statistical
sampling of observed behaviors determines their
acquisition~\cite{hurford89,oliphant_batali,Nowak1999,NowakKrak1999}. The
second emphasizes the functional and inferential nature of
conventional communication, the scaffolding role of the speaker, the
restrictive power of the joint attention frame set up in the shared
context, and the importance of pragmatic feedback in language
interaction.  Here we adopt the reinforcement learning approach as
in~\cite{Hutchins95,steels1995,lenaerts2005}.

In this paper we shall discuss a recently introduced
model~\cite{baronchelli_ng_first}, inspired by one of the first
language game models known as the Naming Game~\cite{steels1995}. It is
able to account for the emergence of a shared set of conventions in a
population of agents. Central control or co-ordination are absent, and
agents perform only pairwise interactions following straightforward
rules. Indeed, due to the simplicity of the interaction scheme, the
dynamics of the model can be studied both with massive simulations and
analytical approaches. By doing so we import a pre-existing model into
the statistical mechanics context (as opposed to the reverse which is
often the case).

In past work, sociocultural investigations largely focused on
computational issues and the application for emergent communication in
software agents or physical robots~\cite{steels03}, resulting in a
lack of quantitative investigations. For instance, we shall discuss in
detail later how the main features of the process leading the
population to a final convergence state scale with the population
size, whereas earlier work has concentrated on studying very small
populations~\cite{smith_iterated}. The price to pay for quantitative
comprehension is a reduction in the number of aspects of the phenomena
we can treat. Thus, the agent architectures we shall describe are
indeed very basic and stylized, and are much too simple compared to
the cognitive mechanisms humans employ, but on the other hand they
allow us to study much more clearly what is crucial to obtain the
desired global co-ordination based on only local interaction. The
present paper shows that the crucial features are in fact simple and
we consider this to be one of our major contributions. Despite
simplifying the original Naming Game~\cite{steels1995}, we retained
however its most important properties so that the interaction scheme
could still be ported to real world robots or be used to explain the
behavior of biological agents.

The paper is organized as follows.  In Sec.~\ref{s:ng_rules} we
present the Naming Game model and discuss its basic
phenomenology. Sec.~\ref{s:system_size} is devoted to the study of the
role of population size. We investigate the scaling relations of some
important quantities and provide analytical arguments to derive the
relevant exponents. In Sec.~\ref{s:approach_to_conv} we look in more
detail at the mechanisms that give rise to convergence, deepening the
analysis presented in~\cite{baronchelli_ng_first}. In particular, we
identify and explain the presence of a hidden timescale that governs
the transition to the final consensus state. In
Sec.~\ref{s:single_games} we focus on the relation between single
simulation runs and averaged quantities, while in
Sec.~\ref{s:conv_word} we investigate the properties of the consensus
word. We then analyze, in Sec.~\ref{s:2words}, a controlled case that
sheds light on the nature of the symmetry breaking process leading to
lexical convergence.  Finally, in Sec.~\ref{s:ng_discussion}, we
discuss the most relevant features of the model and present some
conclusions concerning particularly its connections with the fields of
Opinion Dynamics on one hand and Artificial Intelligence on the other.

\section{The model}

\subsection{Naming Game}\label{s:ng_rules}

We present here the version of the Naming Game introduced
in~\cite{baronchelli_ng_first} (see also~\cite{baronchelli_thesis} for
a comprehensive analysis of the model). The game is played by a
population of $N$ agents in pairwise interactions. As a side effect of
a game, agents {\em negotiate} conventions, i.e., associations between
forms (names) and meanings (for example individuals in the world), and
it is obviously desirable that a global consensus emerges. Because
different agents can each independently invent a different name for
the same meaning, synonymy (one meaning many words) is
unavoidable. However we do not consider here the possibility of
homonymy (one word many meanings). In the invention process, in fact,
we consider the situation where the number of possibly invented words
is so huge that the probability that two players will ever invent the
same word at two different times for two different meanings is
practically negligible. This means that the dynamics of the
inventories associated to different meanings are completely
independent and the number of meanngs becomes a trivial parameter of
the model.  As a consequence we can reduce, without loss of
generality, the environment as composed by one single meaning and
focus on how a population can establish a convention for expressing
that meaning. In a generalized Naming Game, homonymy is not always an
unstable feature and its survival depends in general on the size of
the meaning and signal spaces~\cite{gosti07}. Homonymy becomes crucial
if, during a conversation, agents do not get precise feedback about
the meaning. If there is more than one possible meaning compatible
with the current situation (for example if the word expresses a
category but we do not know which one) then homonymy would be
unavoidable. This is not the case for the Naming Game while it becomes
crucial for the so-called Guessing~\cite{steels_connection} and
Category Game~\cite{puglisi07}.

\begin{figure}[t]
\begin{center}
\includegraphics*[width=10.5cm]{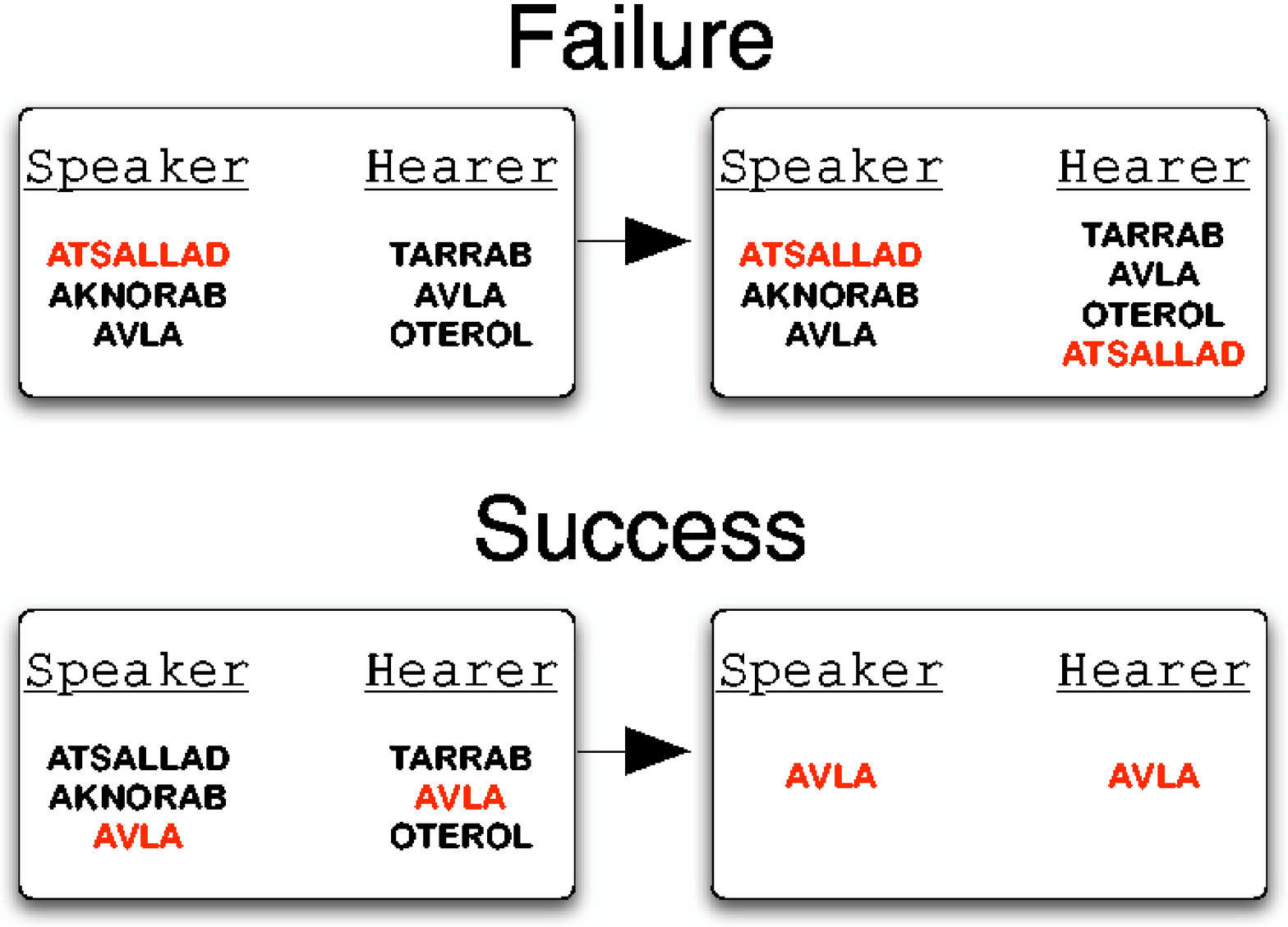}
\caption{{\small {\bf Naming game interaction rules.}  The speaker
    selects randomly one of its names, or invents a new name if its
    inventory is empty (i.e., we are at the beginning of the game). If
    the hearer does not know the uttered name, it simply adds it to
    its inventory, and the interaction is a \ti{failure}. If, on the
    other hand, the hearer recognizes the name, the interaction is a
    \ti{success}, and both agents delete from their inventories all
    their names but the winning one.}}
\label{f:rules}
\end{center}
\end{figure}

The model definition can be summarized as follows. We consider an
environment composed by one single object to be named, the extension
to many different objects being trivial if one neglects homonymy. Each
individual is described by its \ti{inventory}, i.e., a set of
form-meaning pairs (in this case only names competing to name the
unique object)) which is empty at the beginning of the game ($t=0$)
and evolves dynamically in time. At each time step ($t=1,2,..$) two
agents are randomly selected and interact: one of them plays the role
of {\em speaker}, the other one that of {\em hearer}. The interactions
obey the following rules (Fig.~\ref{f:rules}):

\begin{itemize}
\item The speaker transmits a name to the hearer.  If its inventory is
  empty, the speaker invents a new name, otherwise it selects randomly
  one of the names it knows;
  
\item If the hearer has the uttered name in his inventory, the game is
  a \underline{success}, and both agents delete all their names, but
  the winning one;
  
\item If the hearer does not know the uttered name, the game is a
  \underline{failure}, and the hearer inserts the name in its
  inventory.
\end{itemize}

Another important assumption of the model is that two agents are
\textit{randomly} selected at each time step. This means that each
agent in principle can talk to anybody else, i.e., that the population
is completely unstructured (homogeneous mixing assumption). The role
of different agent topologies has been discussed extensively
elsewhere~\cite{ke04-->07,baronchelli_ng_lowdim,dallasta_ng_smallworld,dallasta_ng_nets,dallasta_ng_micro,barrat_chaos_2007,baronchelli_thesis}. A
generalized model of the Naming Game has also been proposed, in which
agents do not update their inventories deterministically after a
success, but rather do that according to a certain
probability~\cite{baronchelli_ng_trans}.  Generalized models exhibit
interesting phenomenologies, including a non-equilibrium phase
transition, but we do not consider them here.

Finally, it is worth stressing that the random selection rule adopted
by the speaker to select the word to be transmitted, and the absence
of weights to be associated with words, expressly violate the
fundamental ingredients of earlier models~\cite{steels_connection}.
Indeed, as we are going to show, they turn out to be unnecessary.

\begin{figure}[t]
\begin{center}
\includegraphics*[width=10.5cm,angle=270]{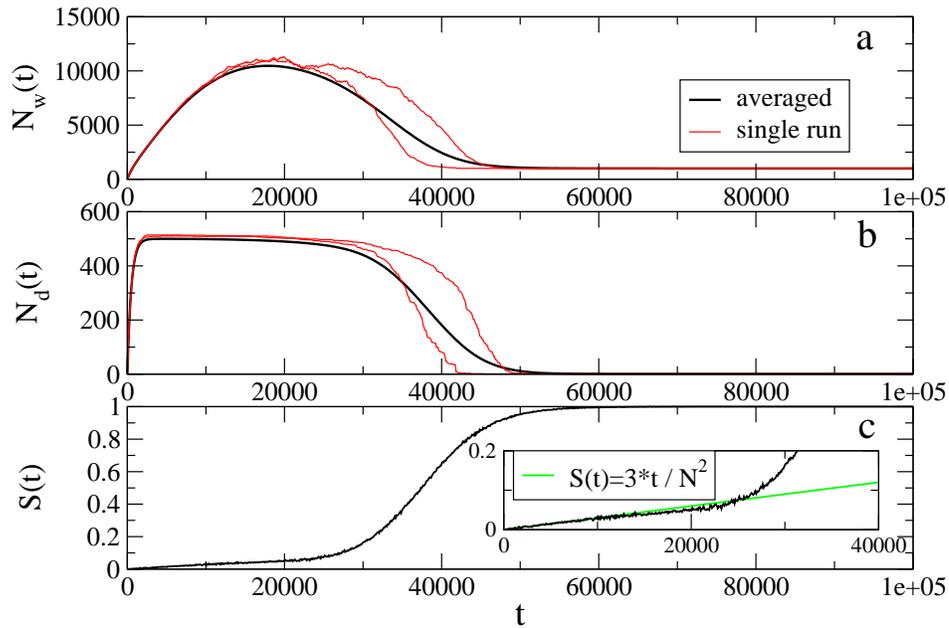}
\caption{{\small {\bf Basic global quantities.} a) Total Number of
    names present in the system, $N_w(t)$; b) Number of different
    names, $N_d(t)$; c) Success rate $S(t)$, i.e., probability of
    observing a successful interaction at a time $t$.  The inset shows
    the linear behavior of $S(t)$ at small times. All curves concern a
    population of $N=10^3$ agents. The system reaches the final
    absorbing state, described by $N_w(t)=N, \;\; N_d(t)=1$ and
    $S(t)=1$, in which a global agreement on the form (name) to assign
    to the meaning (individual object) has been reached.}}
\label{f:classic}
\end{center}
\end{figure}

\subsection{Basic phenomenology}

The most basic quantities describing the state of the population at a
given time $t$ are: the total number of names present in the system,
$N_w(t)$, the number of different names known by agents, $N_d(t)$, and
the success rate, i.e.  the probability of observing a successful
interaction at a given time, $S(t)$.  In Figure~\ref{f:classic} we
report data concerning a population of $N=10^3$ agents. The process
starts with a trivial transient in which agents invent new names. It
follows a longer period of time where the $N/2$ (on average) different
names are exchanged after unsuccessful interactions.  The probability
of a success taking place at this time is indeed very small ($S(t)
\simeq 0$) since each agent knows only a few different names. As a
consequence, the total number of names grows, while the number of
different names remains constant. However, agents keep correlating
their inventories so that at a certain point the probability of a
successful interaction ceases to be negligible. As fruitful
interactions become more frequent the total number of names at first
reduces its growth and then starts to decrease, so that the $N_w(t)$
curve presents a well identified peak. Moreover, after a while, some
names start disappearing from the system. The process evolves with an
abrupt increase in the success rate, with a curve $S(t)$ which
exhibits a characteristic {\em S-shaped}~behavior, and a further
reduction in the numbers of both total and different names. Finally,
the dynamics ends when all agents have the same unique name and the
system is in the desired convergence state. It is worth noting that
the developed communication system is not only \ti{effective} (each
agent understands all the others), but also \ti{efficient} (no memory
is wasted in the final state).

From the inset of Figure~\ref{f:classic} it is also clear that the
$S(t)$ curve exhibits a linear behavior at the beginning of the
process: $S(t) \sim t/N^2$.  This can be understood noting that, at
early stages, most successful interactions involve agents which have
already met in previous games. Thus the probability of success is
proportional to the ratio between the number of couples that have
interacted before time $t$, whose order is $O(t)$, and the total
number of possible pairs, $N(N-1)/2$. The linear growth ends in
correspondence with the peak of the $N_w$ curve, where it holds $S(t)
\sim 1/N^{0.5}$, and the success rate curve exhibits a bending
afterward, slowing down its growth till a sudden burst that
corresponds to convergence.

\section{The role of system size}\label{s:system_size}

\subsection{Scaling relations}
\label{sec:scaling_rel}

\begin{figure}[t]
\begin{center}
  \resizebox{115mm}{!}{\includegraphics*{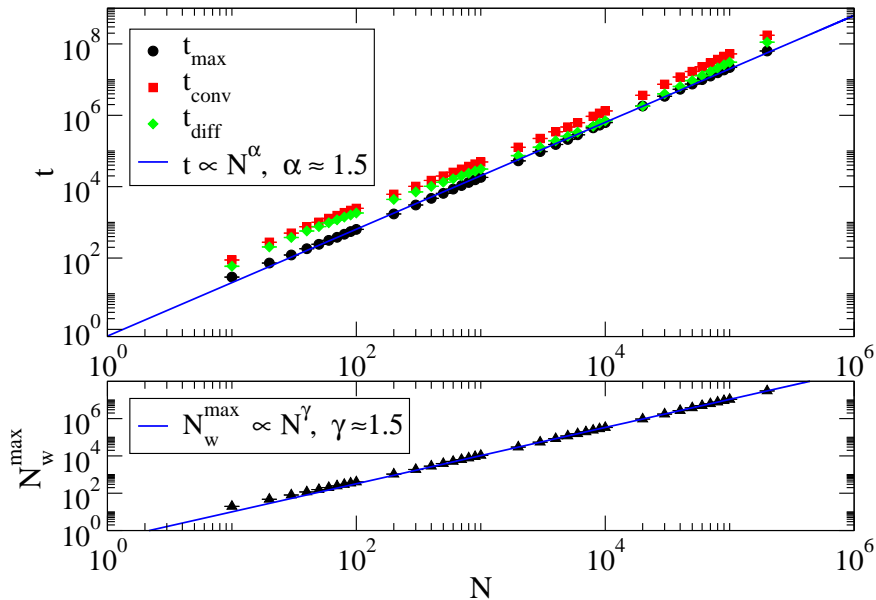}}
\caption{{\small {\bf Scaling with the population size $N$}. In the
    upper graph the scaling of the peak and convergence time,
    $t_{max}$ and $t_{conv}$, is reported, along with their
    difference, $t_{diff}$. All curves scale with the power law
    $N^{1.5}$. Note that $t_{conv}$ and $t_{diff}$ scaling curves
    present characteristic log-periodic oscillations (see
    Sec.~\ref{s:ng_rescaling}). The lower curve shows that the maximum
    number of words (peak height, $N_w^{max} = N_w(t_{max})$) obeys
    the same power law scaling.}}
\label{f:scaling_mf}
\end{center}
\end{figure}

A crucial question concerns the role played by the system size $N$.
In particular, two fundamental aspects depend on $N$. The first is the
time needed by the population to reach the final state, which we shall
call the convergence time $t_{conv}$. The second concerns the
cognitive effort in terms of memory required by each agent in
achieving this dynamics. This reaches its maximum in correspondence of
the peak of the $N_w(t)$ curve. Figure~\ref{f:scaling_mf} shows
scaling of the convergence time $t_{conv}$, and the time and height of
the peak of $N_w(t)$, namely $t_{max}$ and $N_w^{max} \doteq
N_w(t_{max})$. The difference time $(t_{conv} - t_{max})$ is also
plotted. It turns out that all these quantities follow power law
behaviors: $t_{max} \sim N^{\alpha}$, $t_{conv} \sim N^{\beta}$,
$N_w^{max} \sim N^{\gamma}$ and $t_{diff}=(t_{conv}-t_{max}) \sim
N^{\delta}$, with exponents $\alpha \approx \beta \approx \gamma
\approx \delta \approx 1.5$.

The values for $\alpha$ and $\gamma$ can be understood through simple
analytical arguments. Indeed, assume that, when the total number of
words is close to the maximum, each agent has on average $cN^a$ words,
so that it holds $\alpha=a+1$.  If we assume also that the
distribution of different words in the inventories is uniform, the
probability for the speaker to play a given word is $1/(cN^a)$, while
the probability that the hearer knows that word is $2cN^a/N$ (where
$N/2$ is the number of different words present in the system). The
equation for the evolution of the number of words then reads:

\be
\frac{dN_w(t)}{dt} \propto \frac{1}{c N^{a}} \left(1- \frac{2 c
N^{a}}{N}\right) - \frac{1}{c N^{a}} \frac{2 c N^{a}}{N} 2 c
N^{a} 
\ee

\noindent where the first term is related to unsuccessful interactions
(which increase $N_w$ by one unit), while the second one to successful
ones (which decrease $N_w$ by $2cN^a$). At the maximum
${dN_w(t_{max})}/{dt}=0$, so that, in the thermodynamic limit $N
\rightarrow \infty$, the only possible value for the exponent is
$a=1/2$ which implies $\alpha=3/2$ in perfect agreement with data from
simulations.

For the exponent $\gamma$ the procedure is analogous, but we have to use the
linear behavior of the success rate and the relation $a=1/2$ we have just
obtained. The equation for $N_w(t)$ now can be written as:

\be \frac{dN_w(t)}{dt} \propto \frac{1}{c N^{1/2}} \left(1-
  \frac{ct}{N^2}\right) - \frac{1}{c N^{1/2}} \frac{ct}{N^2} 2 c
N^{1/2} \ .  \ee

\noindent If we impose ${dN_w(t)}/{dt}=0$, we find that the time of the
maximum has to scale with the right exponent $\gamma=3/2$ in the thermodynamic
limit.

\begin{figure}[!t]
\begin{center}
  \includegraphics*[width=10.5cm]{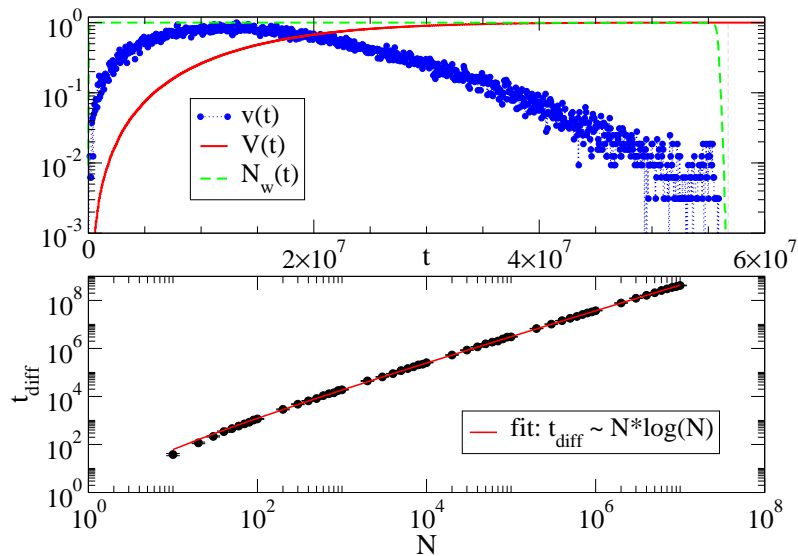}
  \caption{{\small {\bf Evidences supporting the argument for the
        $\beta$ exponent.} Top: $v(t)$ is the (non normalized)
      histogram of the times at which agents play their first
      successful interaction, while $V(t)$ is the cumulative curve. It
      is clear that up to a time very close to convergence there are
      still agents that have never won. Thus, the investigation of the
      first time in which $V(t)=1$ provides a good estimate of
      $t_{conv}$. Data refer to a single run for a population of
      $N=10^5$ agents. The $N_d(t)$ curve is also plotted, for
      reference, while the vertical dashed grey line indicates
      convergence time. Bottom: scaling of $t_{diff}$ with $N$ for a
      system in which, at the beginning of the process, half of the
      population knows word $A$ and the other half word $B$. Thus,
      $N_d(t=0)=2$ and invention is eliminated. Experimental points
      are well fitted by $t_{diff} \sim N \log N$, as predicted by our
      argument (see text). A fit of the form $t_{diff} \sim
      N^{\delta}$, on the other hand, turns out to be less accurate
      (data not shown). }}
\label{f:all_win}
\end{center}
\end{figure}

The exponent for the convergence time, $\beta$, deserves a more
articulate discussion, and we can only provide a more \ti{naive}
argument, even though well supported by evidence from numerical
simulations.  We concentrate on the scaling of the interval of time
separating the peak of $N_w(t)$ and the convergence, i.e.,
$t_{diff}=(t_{conv}-t_{max}) \sim t^{\delta} \sim N^{1.5}$, since we
already have an argument for the time of the peak of the total number
of words $t_{max}$.  $t_{diff}$ is the time span required by the
system to get rid of all the words but the one which survives in the
final state. The problem cast in such a way, we argue that a crucial
parameter is the maximum number of words the system stores at the
beginning of the elimination phase.
 
If we adopt the mean field assumption that at $t=t_{max}$ each agent
has on average $N_w^{max}/N \sim \sqrt{N}$ words
(see~\refcite{dallasta_ng_micro} for a detailed discussion of such a
mean field approximation), we see that, by definition, in the interval
$t_{diff}$, each agent must have won at least once. This is a
necessary condition to have convergence, and it is interesting to
investigate the timescale over which this happens. Assuming that
$\overline{N}$ is the number of agents who did not yet have a
successful interaction at time $t$, we have:
 
\be
\overline{N} = N(1 - p_s p_w)^t
\label{e:scaling_tconv}
\ee

\noindent where $p_s=1/N$ is the probability to randomly select an
agent and $p_w=S(t)$ is the probability of a success. The latter is
$O(1/ N^{0.5})$ at $t_{max}$, and stays around that value for a quite
long time span afterward. Indeed, as we have seen, the success rate
$S(t)$ grows linearly till the peak, where $S(t) = ct_{max}/N^2 \sim
1/N^{0.5}$, and exhibits a bending afterward, before the final jump to
$S(t)=1$ (Fig.~\ref{f:classic}). If we insert the estimates of $p_s$
and $p_w$ in eq.~(\ref{e:scaling_tconv}), and we require the number of
agents who have not yet had a successful interaction to be finite just
before the convergence, i.e., $\overline{N} (t_{conv}) \sim O(1)$, we
obtain $t_{diff} \sim N^{3/2} \log N$. Thus, the leading term of the
difference time $t_{diff} \sim N^{1.5}$ is correctly recovered, and
the necessary condition $\overline{N} (t_{conv}) \sim O(1)$ turns out
to be also sufficient. The possible presence of the logarithmic
correction, on the other hand, cannot be appreciated in simulations
due also to logarithmic oscillations in the $t_{diff}$ curve (see
following Sec.~\ref{s:ng_rescaling}). Finally, it is worth noting that
the $S(t) \sim 1/N^{0.5}$ behavior can be understood also assuming
that at the peak of $N_w(t)$ each agent has $O(N^{0.5})$ words (mean
field assumption), and that the average number of words in common
between two inventories is $O(1)$ (as confirmed by numerical
simulations shown in Fig.~\ref{f:overlap}).

We can test the hypothesis behind the above argument in two
ways. First of all we can investigate the distribution $v(t)$ of the
times at which agents perform their first successful interaction.
Remarkably, Fig.~\ref{f:all_win} (top) shows that this distribution
extends approximately up to $t_{conv}$, so that the time $t^*$, at
which $V(t) \equiv \int_0^{t^*} v(t)=1$, turns out to provide a good
estimate for $t_{conv}$. Then, we can validate our approach studying a
controlled case. Consider a simplified situation in which each agent
starts the usual Naming Game knowing one of only two possible words,
say $A$ and $B$. Invention is then prevented, and for the peak of
$N_w(t)$ it holds $N_w^{max} \sim N$. Noting that in this case we have
$S(t_{max}) \sim O(1)$, and substituting this value in
eq.~(\ref{e:scaling_tconv}), we obtain that $t_{diff} \sim N \log N$.
Indeed, this prediction is confirmed by simulations also for what
concerns the logarithmic correction (Fig.~\ref{f:all_win} (bottom)),
and our approach is supported by a second validation.

\subsection{Rescaling curves}
\label{s:ng_rescaling}

\begin{figure}[t]
\begin{center}
  \includegraphics*[width=10.5cm]{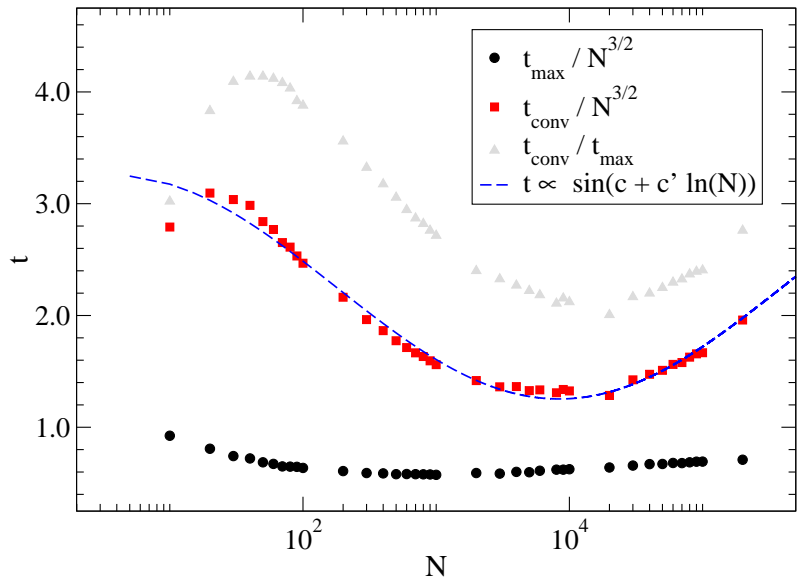}
\caption{{\small {\bf Log-periodic oscillations for convergence times.} Rescaled
values of $t_{conv}$ and $t_{max}$ are plotted along with their ratio. The
rescaled convergence times exhibit global oscillations that are well fitted by
the function $t \propto \sin (c+c' \ln(N))$, where $c$ and $c'$ are constants
whose values are $c \approx 1.0$ and $c' \approx 0.4$.  }}
\label{f:scaling_oscilla}
\end{center}
\end{figure}

Since we know that the characteristic time required by the system to
reach convergence scales as $N^{1.5}$ we would expect a transformation
of the form $t \rightarrow t/N^{3/2}$ to yield a collapse of the
global-quantity curves, such as $S(t)$ or $N_w(t)$, relative to
systems of different sizes.  However this does not happen.

The first reason is that the curve of the scaling of the convergence
time with $N$ does follow a $N^{3/2}$ trend, but presents a peculiar,
seemingly oscillatory, behavior in logarithmic scale. This is already
visible from Figure~\ref{f:scaling_mf}, but is clearer in
Figure~\ref{f:scaling_oscilla}, where it is shown that the curve
$t_{conv}/N^{3/2}$ is well fitted by a function of the type $t \propto
\sin (c+c' \ln(N))$, where $c$ and $c'$ are constants\footnote{It must
  be noted that, since the supposed oscillations should happen on
  logarithmic scale, it is hard to obtain data able to confirm their
  actual oscillatory behavior. Thus, the fit proposed here must be
  intended only as a possible suggestion on the true behavior of the
  irregularities of the $t_{conv}$ scaling curve.}. The same figure
also shows that such oscillations are absent, or at are least very
reduced, in the curve of peak times, $t_{max}$.

\begin{figure}[t]
\begin{center}
\includegraphics*[width=10.5cm]{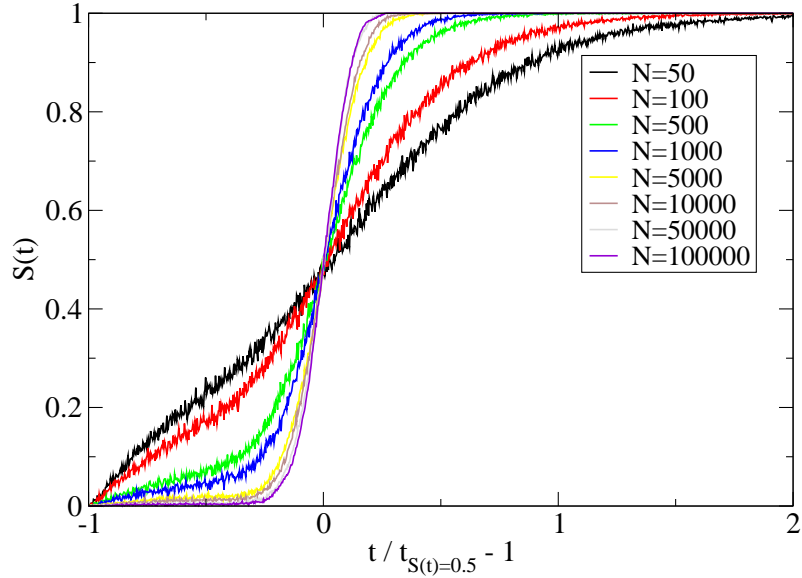}
\caption{{\small {\bf Rescaling of the success rate curves.} Curves
    relative to different system sizes show different qualitative
    behavior if time is rescaled as
    $t~\rightarrow~t / t_{S(t)=0.5} - 1$, where $t_{S(t)=0.5}
    \sim N^{3/2}$. Indeed, on this timescale, the transition between
    the initial disordered state and the final ordered one where $S(t)
    \approx 1$ (i.e., the disorder-order transition) becomes steeper
    and steeper as $N$ grows.}}
\label{f:scaling_pwin_1}
\end{center}
\end{figure}

The deviations of the convergence time scaling curve from a pure power
law have the effect of scattering rescaled curves, thus preventing any
possible collapse.  An easy solution to this problem is that of
rescaling according to intrinsic features of each curve. In
Figure~\ref{f:scaling_pwin_1}, we have rescaled success rate $S(t)$
curves following the transformation $t \rightarrow t/t_{S(t)=0.5} -
1$, where $t_{S(t)=0.5}$ is the time in which the considered curve
reaches the value $0.5$ (with $t_{S(t)=0.5} \sim N^{1.5}$, not shown).
Interestingly we note that the curves still do not collapse. In
particular, the transition between a disordered state in which there
is almost no communication between agents ($S(t) \approx 0$), to the
final ordered state in which most interactions are successful ($S(t)
\approx 1$) becomes steeper and steeper as $N$ becomes
larger~\cite{baronchelli_ng_first}. In other words, it is clear that
the \ti{shape} of the curves changes when we observe them on our
rescaled timescale.

\begin{figure}[t]
\begin{center}
\includegraphics*[width=10.5cm]{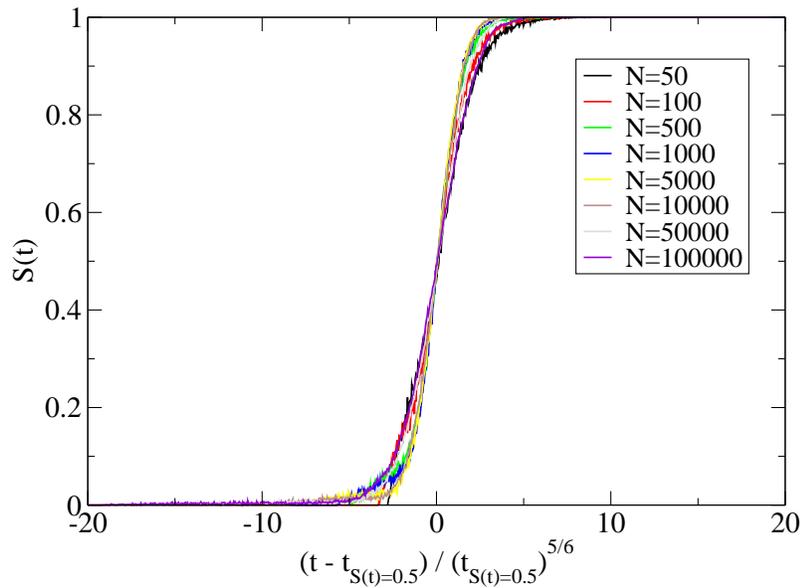}
\caption{{\small {\bf Collapse of the success rate curves.} The time
    rescaling transformation $t \rightarrow
    (t-t_{S(t)=0.5}) / t_{S(t)=0.5}^{5/6}$ makes the different
    $S(t)$ curves collapse. Since the time at which the success rate
    is equal to $0.5$ scales as $N^{3/2}$ (data not shown), the
    transformation is equivalent to $t \rightarrow
    (t-\alpha N^{3/2})/N^{5/4}$.  The collapse shows that the
    disorder-order transition between an initial disordered state in
    which $S(t) \approx 0$ and an ordered state in which $S(t) \approx
    1$ happens on new timescale $t \sim N^\theta$ with $\theta \approx
    5/4$.  }}
\label{f:scaling_pwin_2}
\end{center}
\end{figure}

Figure~\ref{f:scaling_pwin_1} suggests that the disorder-order
transitions happen on a new timescale $t \sim N^\theta$ with $\theta <
\beta$, so that $N^\theta/t_{conv} \rightarrow 0 \;$ when $\;N
\rightarrow \infty$ and the transition becomes instantaneous, on the
rescaled timescale, in the thermodynamic limit. Indeed this is exactly
the case and, as shown in Figure~\ref{f:scaling_pwin_2}, the value
$\theta = 5/4$ and the transformation $t \rightarrow
{(t-\alpha N^{3/2})}/{N^{5/4}}$ produces a good collapse of the success
rate curves relative to different $N$. In the next section we shall
show how the right value for $\theta$ can be derived with scaling
arguments after a deeper investigation of the model
dynamics~\cite{baronchelli_ng_first}.

\section{The approach to convergence}\label{s:approach_to_conv}

\subsection{The domain of agents}\label{s:domain_agents}

\begin{figure}[t]
\begin{center}
\includegraphics*[width=10.5cm]{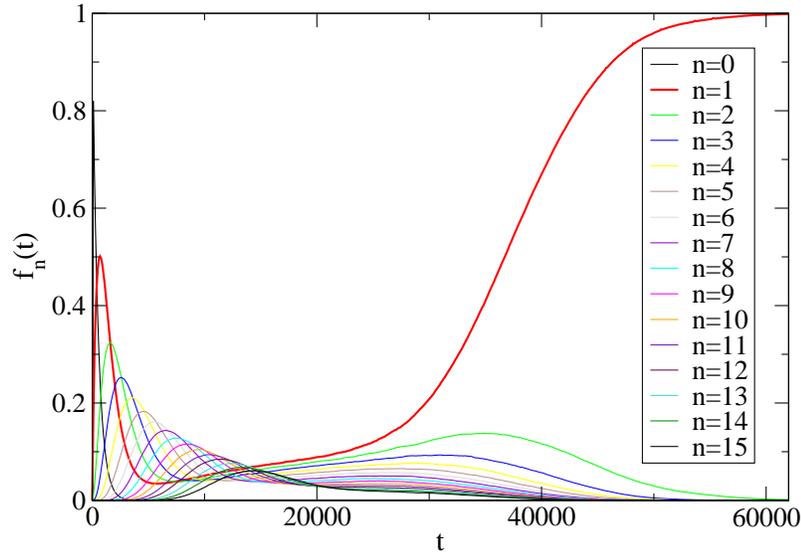}
\caption{{\small {\bf Evolution in time of inventory sizes $n$ ($n=1
      \ldots 15$).} $f_n(t)$ is the fraction of agents whose inventory
    size is $n$ at time $t$. In the right part, $f_n(t)$ decreases
    with increasing $n$. The process ends with all agents having the
    same unique word in their inventory, so that $f_1=1$. Curves
    obtained by averaging $500$ simulation runs on a population of
    $N=10^3$ agents.}}
\label{f:distr_k}
\end{center}
\end{figure}

We have seen that agents at first accumulate a growing number of words
and then, as their interactions become more and more successful,
reduce the size of their inventories till the point in which all of
them know the same unique word.  More quantitatively, the evolution in
time of the fraction of agents $f_n$ with inventory sizes $n$ is shown
in Figure~\ref{f:distr_k}. The curves refer to a population of
$N=10^3$ agents and have been obtained averaging over several
simulation runs. We see that the process starts with a rapid decrease
of $f_0$ and a concomitant increase of the fraction of agents with
larger inventories. After a while, however, successful interactions
produce a new growth in the fraction of agents with small values of
$n$. The process evolves until the point in which all agents have the
same unique word and $f_1=1$.
 
Some of the initial-time regularities of the $f_n$ curves can be
easily described analytically. For instance, it is easy to write
equations for the evolution of the number of species as long as
$S(t)=0$. We have:
  
\bea 
\label{eq:trivial}
{df}_0/dt &=& - f_0 \\
\nonumber {df}_{n > 1}/dt &=& f_{n-1} - f_n \\
\eea
  
\noindent These trivial relations allow to understand some features of
the curves, like the exponential decay of $f_0$, or the fact that, at
early times, each $f_n$ ($n>0$) crosses the correspondent $f_{n-1}$ in
correspondence of its maximum (as can be recovered imposing
${df}_n/dt=0$). However, generalizing eq.~(\ref{eq:trivial}) is not
easy, since, as the dynamics proceeds, one should take into account
the correlations among inventories to estimate the probability of
successful interactions, and the analytical solution of our Naming
Game model is still lacking.

\begin{figure}[t]
\begin{center}
  \includegraphics*[width=10.5cm]{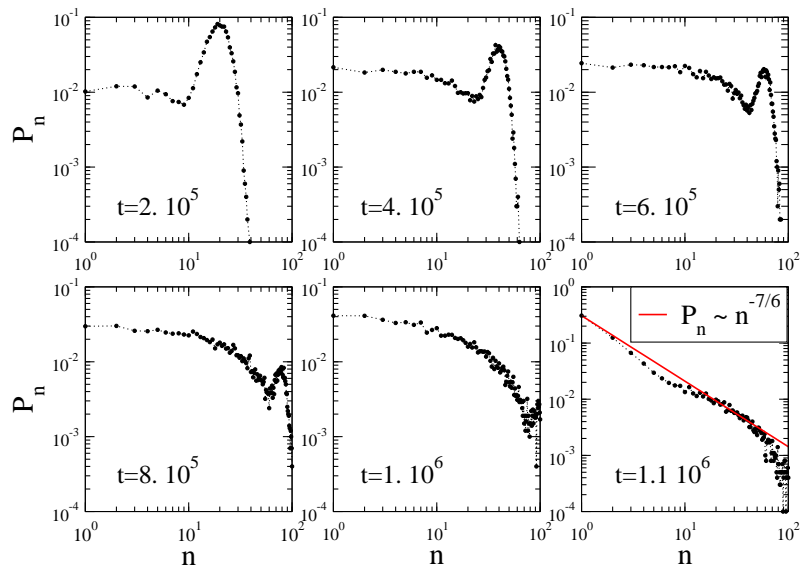}
\caption{{\small {\bf Distribution $P_{n}$ of inventory sizes $n$.} Curves obtained by a single
simulation run for a population of $N=10^4$ agents, for which $t_{max}=6.2
\times 10^5$ and $t_{conv}=1.3 \times 10^6$ time steps. Close to convergence the
distribution is well described by a power law $P_{n} \sim n^{-7/6}$.  }}
\label{f:isto_k}
\end{center}
\end{figure}

More quantitative insights can be obtained looking at the distribution
$P_{n}$ of inventory sizes $n$ at fixed
times~\cite{baronchelli_ng_first}, reported in Figure~\ref{f:isto_k}
for the case $N=10^4$ (see~\cite{dallasta_ng_micro} for a detailed
discussion of the $P_{n}$ behavior in different temporal regions and
different topologies).  We see that in early stages most agents tend
to have large inventories, thus determining a peak in the
distribution. When agents start to understand each other, however, the
peak disappears and large $n$ values keep decreasing. Interestingly,
in correspondence with the jump of the success rate that leads to
convergence, the histogram can be described by a power law
distribution:

\be
P_{n} \sim n^{-{\sigma}} g(n/\sqrt{N})
\label{p(k)}
\ee

\noindent with the cut-off function $g(x) = 1$ for $x << 1$ and $g(x)
= 0$ for $x >> 1$. Numerically it turns out that $ 1 < \sigma < 3/2 $.
To be more precise, in Figure~\ref{f:isto_k} it is shown that the
value $\sigma \approx 7/6$ allows a good fitting of the $P_{n}$ at the
transition, and from simulations it turns out that this is true
irrespectively of the system size.

Finally, it is also worth mentioning that, well before the transition,
the larger number of words in the inventory of a hearer increases
(linearly) the chances of success in a interaction (data not
shown). The number of words known by the speaker, on the other hand,
basically plays no role until the system is close to the transition.
Here, small inventories are likely to contain the most popular word,
thus yielding higher probability of success~\cite{dallasta_ng_micro}.

\subsection{The domain of words}\label{sym_break1}

\begin{figure}[t]
\begin{center}
  \includegraphics*[width=10.5cm]{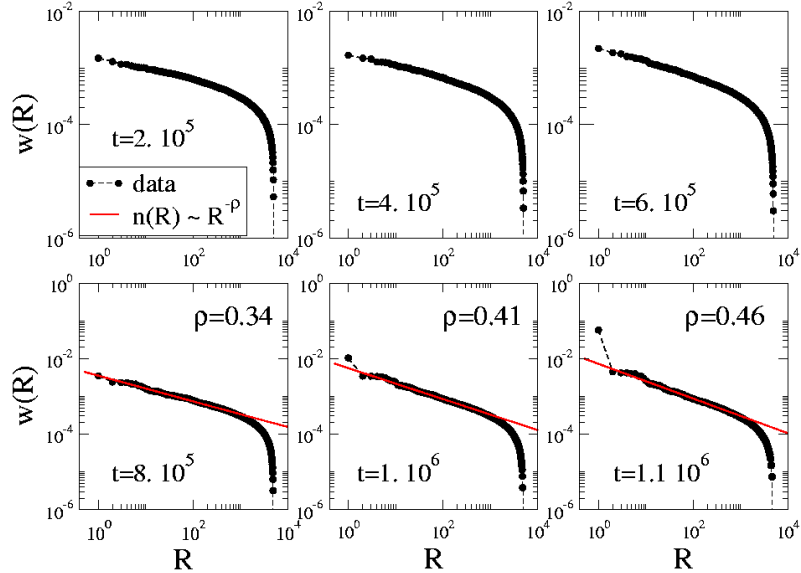}
\caption{{\small {\bf Distribution $w(R)$ of words of rank $R$.} The
    most popular word has rank $R=1$, the second $R=2$, etc. The
    distribution follows a power law behavior $w(R) \sim R^{-\rho}$
    with an exponent that varies in time, while for high ranks it is
    truncated at $R \approx N/2$. Close to the disorder-order
    transition, however, the most diffused word abandons the
    distribution that keeps describing the less popular words. Data
    come from a single simulation run and concern a population of
    $N=10^4$ agents.  }}
\label{f:zipflike}
\end{center}
\end{figure}

While agents negotiate with each others, words compete to
survive~\cite{baronchelli_ng_first}.  In Figure~\ref{f:zipflike} the
rank distribution of words at fixed times is reported. The most
popular word is given rank $1$, the second one $2$ and so on.  The
first part of the distribution is well described by a power law
function, with an exponent that decreases with time. In proximity of
the disorder-order transition, however, the most popular word breaks
the symmetry and abandons the power law behavior, which continues to
describe well the remaining words. More precisely, the global
distribution for the fraction of agents possessing the $R$-ranked
word, $w(R)$, can be described as:

\be 
w(R) = w(1) \delta_{R,1} +
\frac{N_w/N-w(1)}{(1-\rho)((N/2)^{1-\rho}-2^{1-\rho})} R^{-\rho} g(\frac{R}{N/2}),
\label{w(R)}
\ee

\noindent where $\delta$ is the Kronecker delta function
($\delta_{a,b} = 1$ iff $a=b$ and $\delta_{a,b} = 0$ if $a \neq b$)
and the normalization factors are derived imposing that $\int_1^\infty
w(R) dR = N_w/N$~\footnote{We use integrals instead of discrete sums,
  an approximation valid in the limit of large systems.}.

On the other hand from equation (\ref{p(k)}) one gets, by a simple
integration, the relation $N_w/N \sim N^{1-\sigma/2}$ which,
substituted into eq. (\ref{w(R)}), gives:

\be w(R)|_{R > 1} \sim \frac{1}{N^{\sigma/2-\rho}} R^{-\rho}
f(\frac{R}{N/2}).  \ee
 
\noindent It follows that $w(R)|_{R > 1} \rightarrow 0$ as $N
\rightarrow \infty$, so that, in the thermodynamic limit $w(1)\sim
O(1)$, i.e., the number of players with the most popular word, is a
finite fraction of the whole population.
  
\subsection{Network view - The disorder-order transition}  

\begin{figure}[!t]
  \vspace{1.5cm}
\begin{center}
\includegraphics*[width=0.48\textwidth]{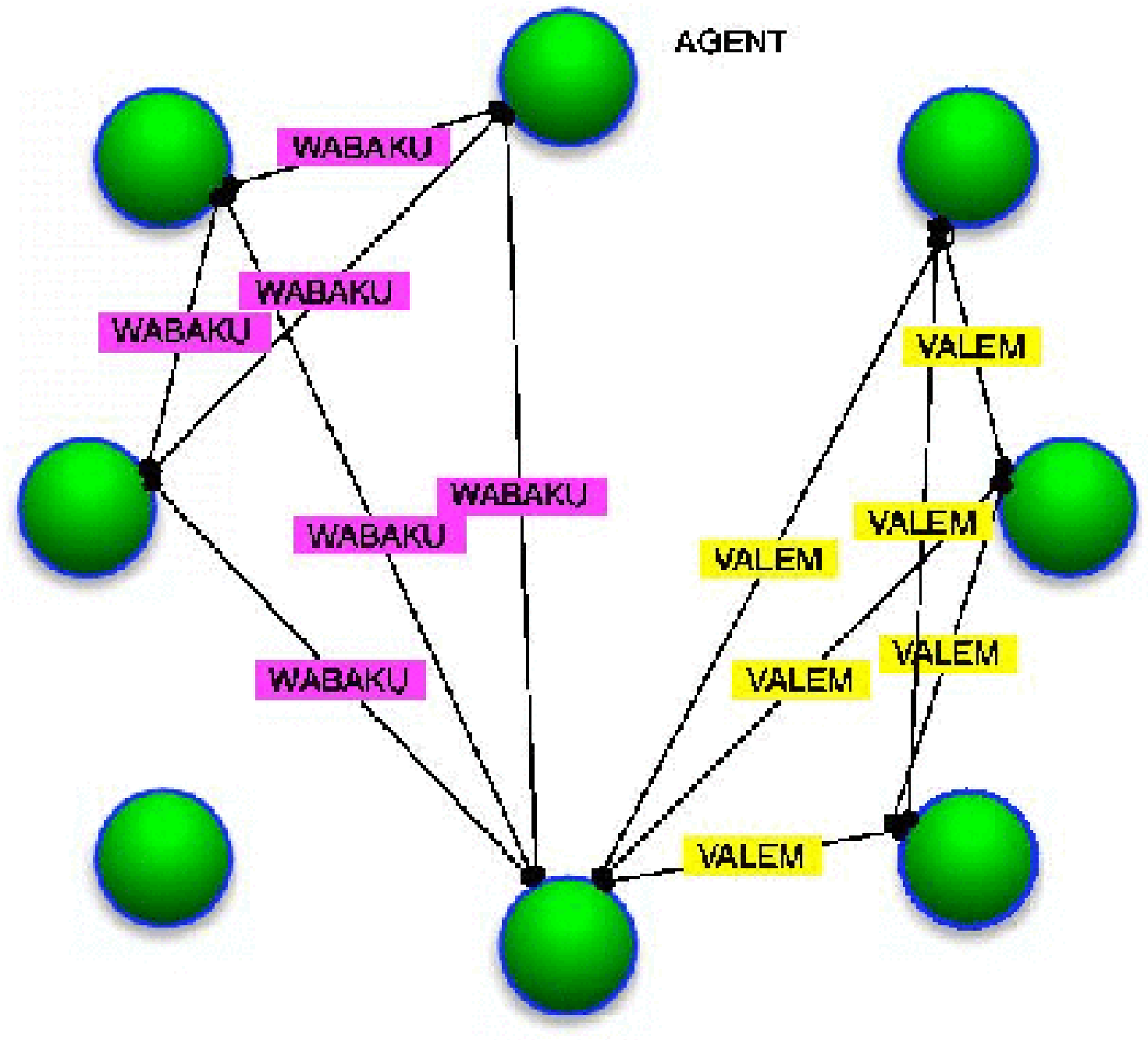} 
\includegraphics*[width=0.48\textwidth]{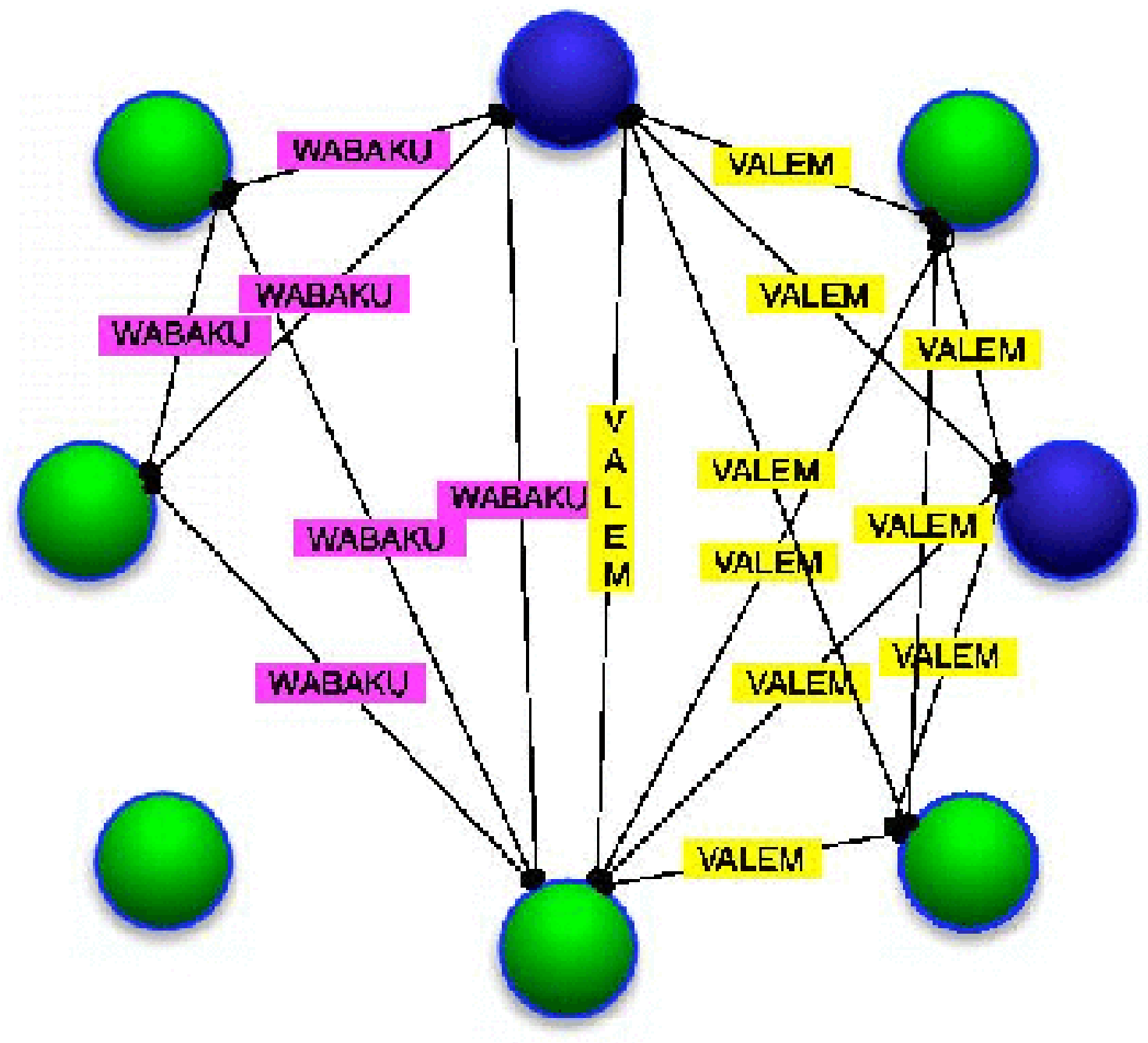} \\
\vspace{0.3cm}
\includegraphics*[width=0.48\textwidth]{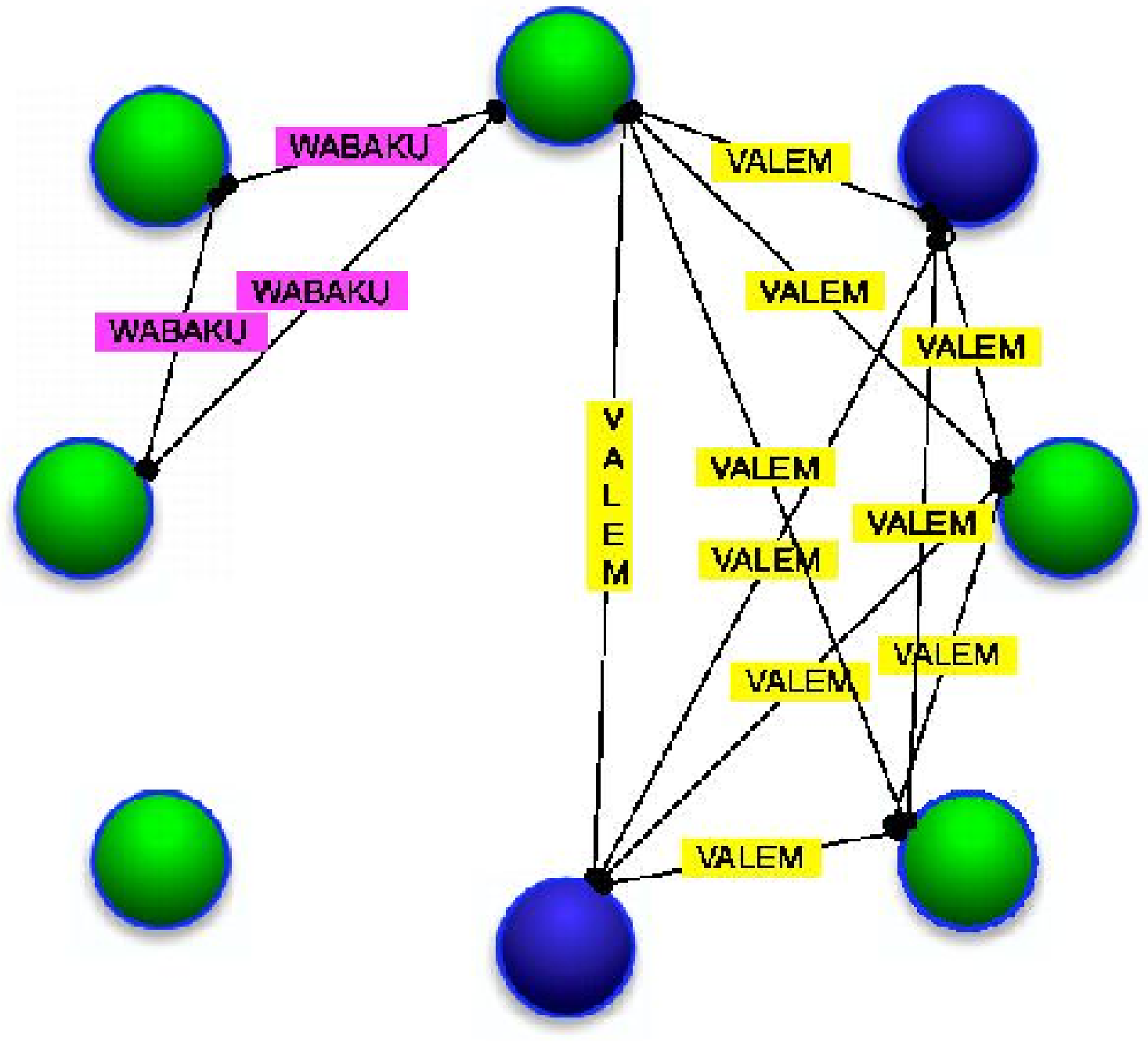} 
\vspace{0.6cm}
\caption{{\small {\bf Agents network dynamics.} Top Left: a link
    between two agents (i.e., nodes) exists every time they have a
    word in common in their inventories, so that multiple links are
    allowed. In this representation, a word corresponds to a fully
    connected (sub)set of agents, i.e., a clique; in Figure, the two
    cliques corresponding to words {\footnotesize WABAKU} and
    {\footnotesize VALEM} are highlighted. Top Right: the two
    highlighted agents have just failed to communicate, so that the
    word {\footnotesize VALEM} has been transmitted to the agent
    placed in the top of the graphical representation. It therefore
    enters into the enlarged clique corresponding to the transmitted
    word {\footnotesize VALEM}. Bottom: the two highlighted agents
    have just succeeded using word {\footnotesize VALEM}. The clique
    corresponding to the used word does not change in any respect, but
    the competing cliques (here that of {\footnotesize WABAKU}) are
    reduced.}} \label{f:net_view}
\end{center}
\end{figure}

We now need a more precise description of the convergence
process~\cite{baronchelli_ng_first}. A profitable approach consists in
mapping the agents in the nodes of a network (see
Figure~\ref{f:net_view}). Two agents are connected by a link each time
that they both know the same word, so that multiple links are
allowed. For example, if $m$ out of the $n$ words known by agent $A$
are present also in the inventory of agent $B$, they will be connected
by $m$ links. In the network, a word is represented by a fully
connected sub-graph, i.e., by a clique, and the final coherent state
corresponds to a fully connected network with all pairs connected by
only one link.  When two players interact, a failure determines the
propagation of a word, while a success can result in the elimination
of a certain number of words competing with the one used. In the
network view, as shown in Figure~\ref{f:net_view}, this translates
into a clique that grows when one of its nodes is represented by a
speaker that takes part in a failure, and is diminished when one (or
two) of its nodes are involved in a successful interaction with a
competing word.

To understand why the disorder-order transition becomes steeper and
steeper, if observed on the right timescale, we must investigate the
dynamics that leads to convergence. If we make the hypothesis that,
when $N$ is large, just before the transition all the agents have the
word that will dominate, the problem reduces to the study of the rate
at which competing words disappear. In different words, the crucial
information is how the number of deleted links in the network, $M_d$,
scales with $N$. It holds:

\be
M_d = \frac{N_w}{N} \int_{2}^{\infty} w^2(R) N dR \sim N^{3 - \frac{3}{2}\sigma}
\label{e:thirdscale}
\ee

\noindent where $\frac{N_w}{N}$ is the average number of words known
by each agent, $w(R)$ is the probability of having a word of rank
$R$, and $w(R)N$ is the number of agents that have that word (i.e.,
the size of the clique). On the other hand, considering the network
structure, eq.~\ref{e:thirdscale} is the product of the average number
of cliques involved in each deletion process [$\frac{N_w}{N}$],
multiplied by an integral stating, in probability, which clique is
involved [$w(R)$] and which is its size [$w(R)N$].  The integral on
$R$ starts from the first deletable word, i.e., the second most
popular, because of the assumption that all the successes are due to
the use of the most popular word.

In our case, for $\sigma \approx 7/6$, we obtain that $M_d \sim
N^{5/4}$. Thus, from equation (\ref{e:thirdscale}), we have that the
ratio $M_d/N^{3/2} \sim N^{-\frac{3}{2}(\sigma-1)}$ goes to zero for
large systems (since $\sigma \approx 7/6$, and in general $\sigma >
1$), and this explains the greater slope, on the system timescale, of
the success rate curves for large populations
(Figure~\ref{f:scaling_pwin_2}).

\subsection{The overlap functional} \label{s:overlap}

We have looked at all the timescales involved in the process leading
the population to the final agreement state. Yet, we have not
investigated \ti{whether} this convergence state is always
reached. Actually, this is the case, and trivial considerations allow
to clarify this point. First of all, it must be noticed that,
according to the interaction rules of the agents, the agreement
condition constitutes the only possible absorbing state of our
model. The proof that convergence is always reached is then
straightforward. Indeed, from any possible state there is always a
non-zero probability to reach an absorbing state in, for instance,
$2(N-1)$ interactions.  For example, a possible sequence is as
follows. A given agent speaks twice with all the other $(N-1)$ agents
using always the same word $A$. After these $2(N-1)$ interactions all
the agents have only the word $A$. Denoting with $p$ the probability
of the sequence of $2(N-1)$ steps, the probability that the system has
not reached an absorbing state after $2(N-1)$ iterations is smaller or
equal to $(1-p)$. Therefore, iterating this procedure, the probability
that, starting from any state, the system has not reached an absorbing
state after $2k(N-1)$ iterations, is smaller than $(1-p)^k$ which
vanishes exponentially with $k$.  The above argument, though being
very simple and general, is exact. However, another perspective to
address the problem of convergence consists in monitoring the lexical
coherence of the system.  To this purpose, we introduce the overlap
functional $O$:

\be
O(t)=\frac{2}{N(N-1)}\sum_{i>j} \frac{|a_i \cap a_j|}{k_i k_j},
\label{e:overlap}
\ee

\noindent where $a_i$ is the $i^{th}$ agent's inventory, whose size is
$k_i$, and $|a_i \cap a_j|$ is the number of words in common between
$a_i$ and $a_j$. The overlap functional is a measure of the lexical
coherence in the system and it is bounded, $O(t) \leq 1$. A the
beginning of the process it is equal to zero, $O(t=0)=0$, while at
convergence it reaches its maximum, $O(t=t_{conv})=1$.

\begin{figure}[!t]
\begin{center}
\includegraphics*[width=10.5cm]{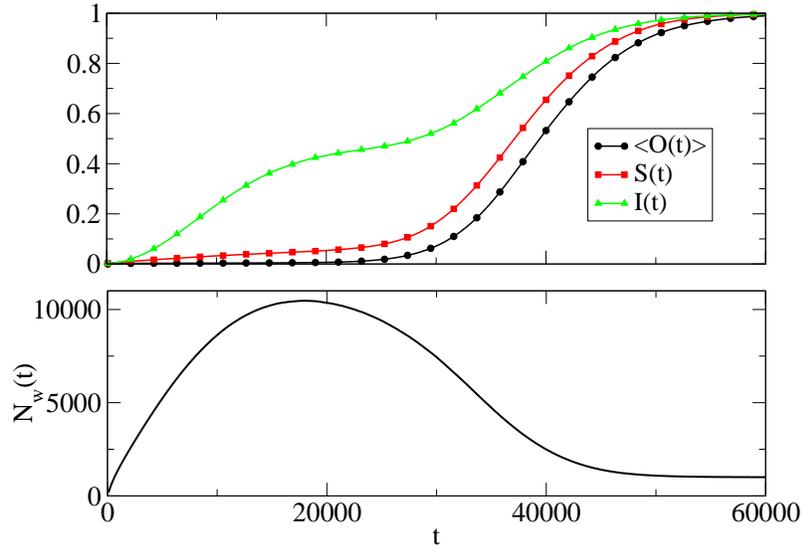} 
\caption{{\small {\bf Overlap functional $O(t)$.} Top: it is shown the
    evolution in time of the overlap functional averaged on $1000$
    simulation runs (for a population of $N=10^3$ agents). Curves for
    the success rate, $S(t)$, and the average intersection between
    inventories, $I(t)$, are also included. By definition, $O(t) \leq
    1$. It is evident that it holds $\la O(t+1) \ra > \la O(t) \ra$,
    which, along with the stronger $\la O(t+1) \ra > O(t) $ valid for
    almost all configurations (not shown), indicate that the system
    will reach the final state of convergence where $O(t)=1$. Bottom:
    The total number of words $N_w(t)$ is plotted for reference. }}
\label{f:overlap}
\end{center}
\end{figure} 

From extensive numerical investigations it turns out that, averaged
over several runs, the functional always grows, i.e., $\la O(t+1) \ra
> \la O(t) \ra $ (see Figure~\ref{f:overlap}). Moreover, looking at
the single realization, this function grows almost always, i.e., $\la
O(t+1) \ra > O(t)$, except for a set a very rare configurations whose
statistical weight appears to be negligible (data not shown). Even if
it is not a proof in a rigorous sense, this monotonicity, combined
with the fact that the functional is bounded, gives a strong
indication that the system will indeed
converge~\cite{baronchelli_ng_first}.

It is also interesting to note that eq.~(\ref{e:overlap}) is very similar 
to the expression for the success rate $S(t)$, which can formally be written as:

\be S(t)=\frac{1}{N(N-1)}\sum_{i>j} \left( \frac{|a_i \cap a_j|}{k_i}
  + \frac{|a_i \cap a_j|}{k_j}, \right) \ee

\noindent where the intersection between two inventories are divided
only by the inventory size of the speaker. Figure~\ref{f:overlap}
shows that these two quantities exhibit a very similar
behavior. However, while the overlap functional is equal to $1$ only
at convergence, this is not true for the success rate: if all agents
had identical inventories of size $n>1$ we would have $S(t)=1$ and
$O(t)=1/n$. For this reason the success rate is not a suitable
functional to prove convergence.

Finally, in Fig.~\ref{f:overlap} we have plotted also the average
intersection between inventories, i.e.

\be
I(t)=\frac{2}{N(N-1)}\sum_{i>j} |a_i \cap a_j|.
\ee 

\noindent Remarkably, it turns out that $I(t)<1$ during all the
process, even if in principle this quantity is not bounded.

\section{Single games}\label{s:single_games}

\begin{figure}[!t]
\begin{center}
\includegraphics*[width=10.5cm]{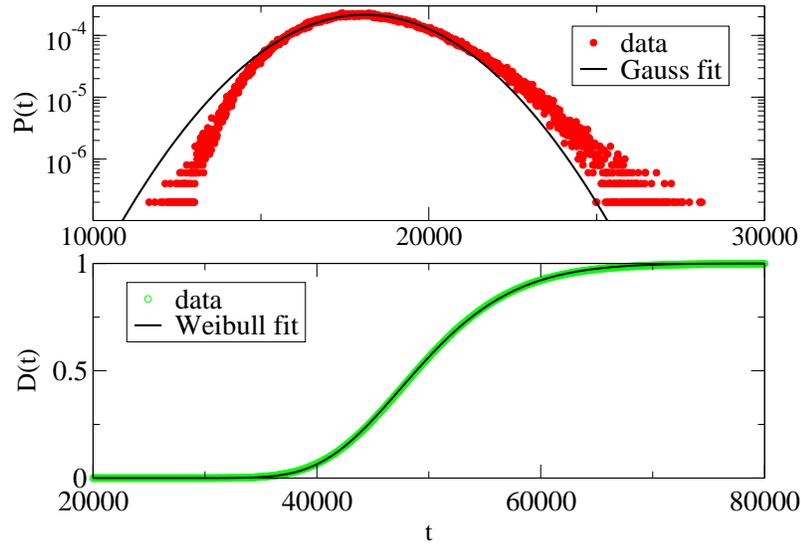}
\caption{{\small {\bf Peak and convergence time distributions.} Top:
    the distribution of the peak times $t_{max}$ clearly deviates from
    Gauss behavior.  Bottom: the cumulative distribution of the
    convergence times $t_{conv}$ is well fitted by a Weibull
    distribution $D(t) = (\exp{(\frac{t-g_0}{g_1}))^{g_2}}$, with fit
    parameters $g_0 \approx 4.9 \times 10^4$, $g_1 \approx 7.9 \times
    10^0$ and $g_2 \approx 9.6 \times 10^4$. The same function
    describes well also the peak time distribution (data not
    shown). Data refers to a population of $N=10^3$ agents and are the
    result of $10^6$ simulation runs.}}
\label{f:extval}
\end{center}
\end{figure}

We know that single realizations have a quite irregular behavior and
can deviate significantly from average curves (Fig.~\ref{f:classic}).
It is therefore interesting to investigate to what extent \ti{average}
times and curves provide a good description of single processes.

In Figure~\ref{f:extval}(top) we have plotted the distribution of peak
times for a population of $N=10^3$ agents. It is clear that data
cannot be fitted by a Gaussian distribution. The same peculiar
behavior is shown also by the distribution of the convergence times
(Fig~\ref{f:extval}(bottom)) and by that of the intervals between the
time of the maximum number of words and the time of convergence (data
not shown). Thus, the non-Gaussian behavior appears to be an intrinsic
feature of the model. In fact, as shown in
Figure~\ref{f:extval}(bottom) for the convergence times, all these
distributions turn out to be well fitted (in their cumulative form) by
an extreme value distribution:

\be
D(t) = \exp{(\frac{t-g_0}{g_1})^{g_2}}
\ee

\noindent where $g_0$, $g_1$ and $g_2$ are fit
parameters~\cite{extval_gumbel,galambos1978}.

\begin{figure}[!t]
\begin{center}
\includegraphics*[width=10.5cm]{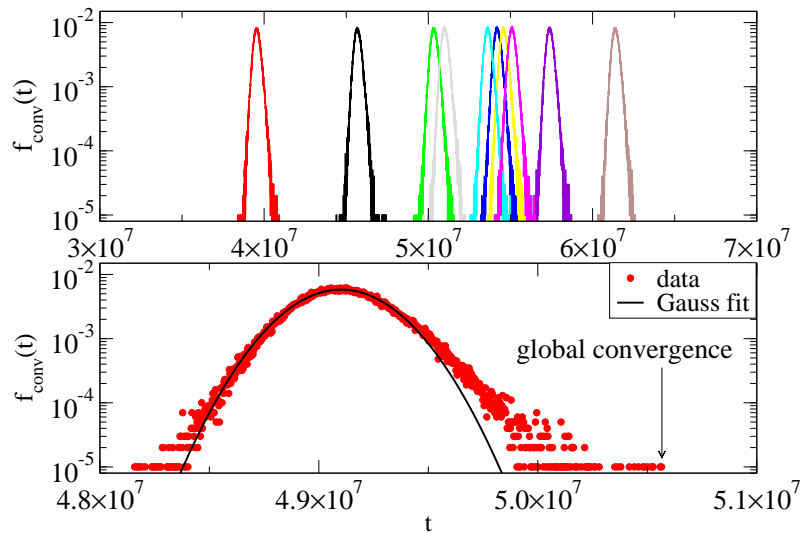}
\caption{{\small {\bf Single agents convergence time distribution.} We
    define the convergence time of a single agent the last time in
    which it had to delete words after a successful interactions;
    $f_{conv}(t)$ is the fraction of agents who reach convergence at
    time $t$. Top: distributions coming from $10$ simulation runs are
    plotted.  It is clear that distributions coming from different
    runs can be non-overlapping, i.e., that the distance between the
    peaks of single curves can be much larger than the average width
    of the same curves (that does not exhibit any strong dependence on
    the single run). Bottom: a single distribution is analyzed,
    showing that it can not be described by a Gauss distribution.  The
    last agent to converge determines the global convergence
    time. Curves are relative to a population of $N=10^5$ agents.  }}
\label{f:extval2}
\end{center}
\end{figure} 

Extreme value distributions originated from the study of the
distribution of the maximum (or minimum) in a large set of independent
and identically distributed set of
variables~\cite{extval_gumbel,galambos1978}. It turns out, however,
that a generalization of these functions including a continuous shape
parameter $a$, known as Gumbel distribution $G_a(x)$, has been
observed in many models ranging from turbulence and equilibrium
critical systems~\cite{extval_turbolence} to non-equilibrium models
related to self-organized criticality~\cite{extval_nonequilibrium}, to
$1/f$ noise~\cite{extval_noise} and many others systems
(see~\cite{extval_bertin} and references therein). The Naming Game
model provides another example.

\begin{figure}[t]
\begin{center}
\includegraphics*[width=10.5cm]{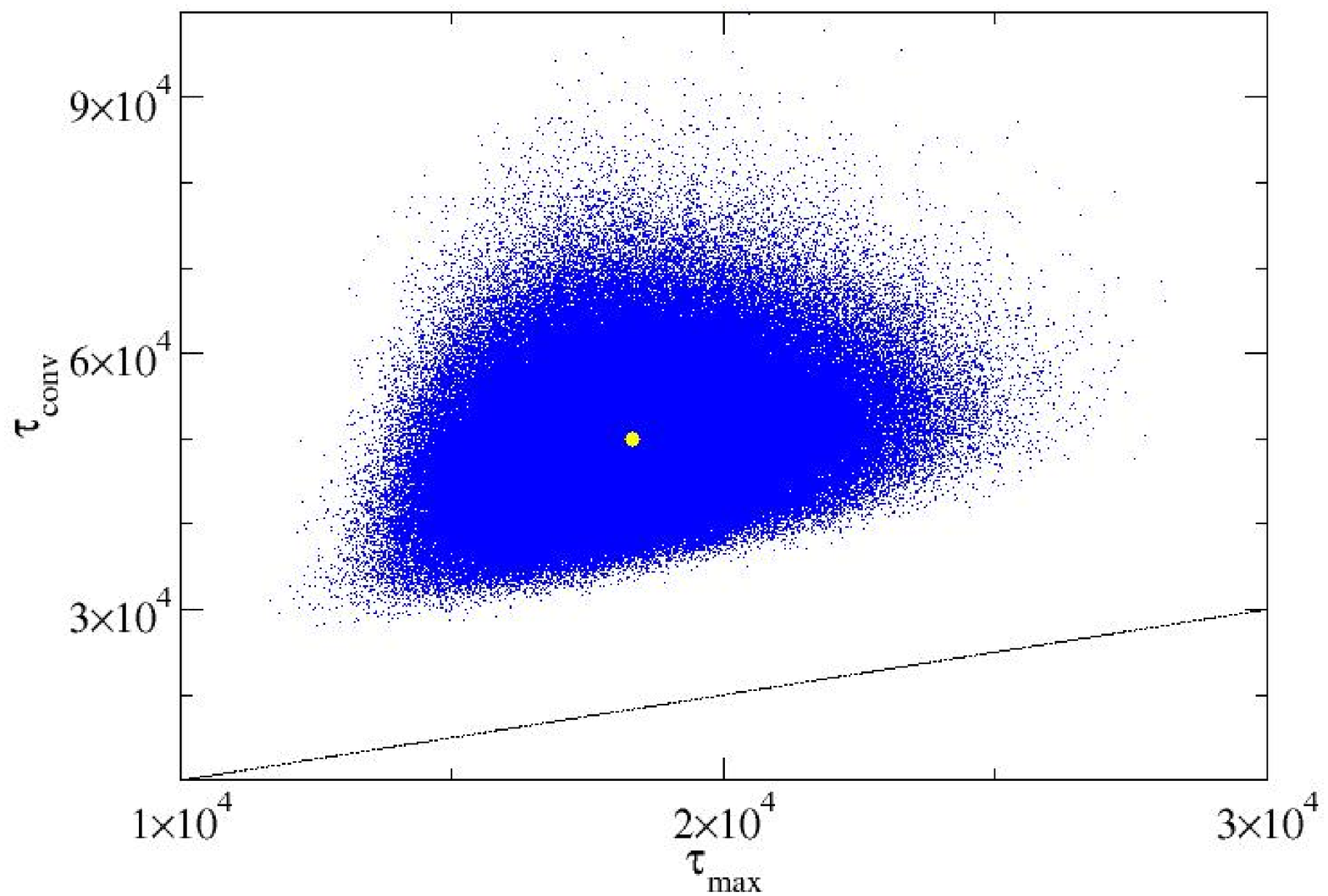}
\caption{{\small {\bf Correlation between peak and convergence times} ($\tau_{max}$ 
$\tau_{conv}$, respectively). Each run is represented by a point in the scatter
plot. The dashed line is $\tau_{conv} =
\tau_{max}$ and therefore no points can lay below it. The average times
$t_{conv}$ and $t_{max}$ are also shown with a clearer (yellow) point at the
center of the distribution (statistical errors are not visible
on the scale of the graph). }}
\label{f:correlation}
\end{center}
\end{figure} 

\begin{figure}[t]
\begin{center}
\includegraphics*[width=10.5cm]{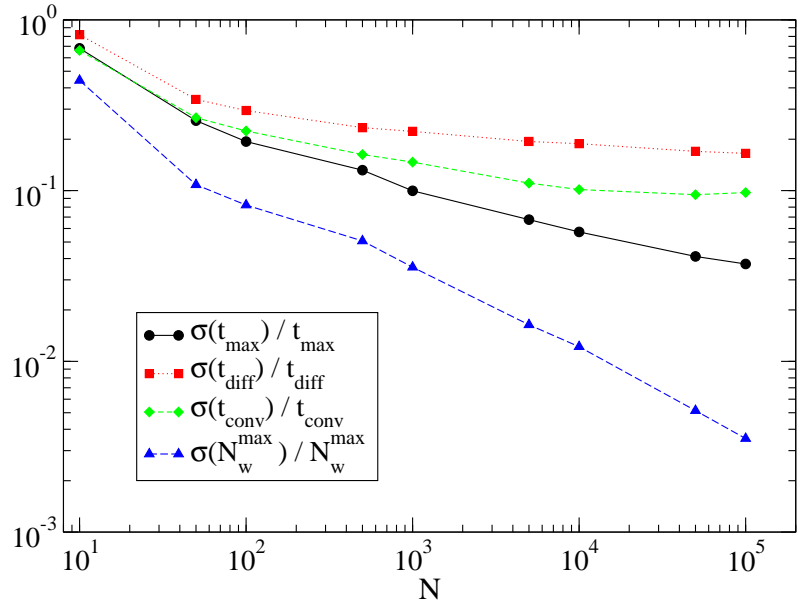}
\caption{{\small {\bf Scaling of the relative standard deviation
      $\sigma(x) / \la x \ra$.} The ratio between the standard
    deviation $\sigma$ and the corresponding (average) quantity is
    plotted as a function of the system size. In all cases the ratio
    decreases slightly, or stays constant, as the population size $N$
    grows. In particular, the decrease is more evident for $N_w^{max}$
    and $t_{max}$, while $t_{conv}$ and $t_{diff}$ curves are almost
    constant for large $N$.  However, data from our simulations are
    not sufficient to conclude whether the Naming Game exhibits
    self-averaging. The standard deviation of $x$ is defined as
    $\sigma(x) = \sqrt{\frac{1}{N_{runs}-1}\sum_{i=1}^{N_{runs}} (x_i
      - \la x \ra)^2}$, $x_i$ is the $i_{th}$ measured value, $\la x
    \ra$ is the average value, and \ $N_{runs}$ is the number of
    simulation runs (here, $N_{runs}=1000 $). }}
\label{f:relative_error}
\end{center}
\end{figure} 

It must be noted, however, that there is no obvious theoretical
explanation of the fact that extreme-value like distributions are
found also in the study of the fluctuations of global quantities. Yet,
in many cases, these distributions are used simply like convenient
fitting functions. Interestingly, it was recently shown that there is
a connection between Gumbel functions and the statistics of global
quantities expressed as sums of non identically distributed random
variables, without the need of invoking extremal
processes~\cite{extval_bertin}. We can therefore argue that there is
not necessarily a hidden extreme value problem in our model. In any
case, a more rigorous explanation of the presence of Gumbel like
distribution is left for future work.

In Figure~\ref{f:extval2}(top) we show 10 single-run distributions of
convergence times. Each curve illustrates the fraction of agents that
converged at a given time in that run, $f_{conv}(t)$. We consider the
single agent convergence time as the last time in which it had to
delete words after a successful interaction.  From Figure it is clear
that the separation between the peaks of two different distributions
can be much larger than the average width of a single curve.  In other
words, we see that the first moment of the distributions strongly
depends on the single realization, while the second one does not. This
information is crucial to interpret the curves shown in
Figure~\ref{f:extval} correctly. In fact, we now know that they are
indeed representative of fluctuations occurring among different runs,
and do not describe simply the behavior of the last converging agent
in a scenario in which most agents always converge, on average, at the
same run-independent time. In Figure~\ref{f:extval2}(bottom) it is
shown that single run curves also deviate from Gauss behavior showing
long tails for large times.

Given these distributions of convergence and peak times, and also that
their difference $t_{diff}$, behaves in the same way, it is
interesting to investigate whether there is any correlation between
these two times. In Figure~\ref{f:correlation} we present a scatter
plot in which the axis indicate $\tau_{conv}$ and $\tau_{max}$,
respectively the convergence and peak times for a single run (so that
$t_{max} = \langle \tau_{max} \rangle$ and $t_{conv} = \langle
\tau_{conv} \rangle$). It is clear that the correlation between this
two times is very feeble. Indeed, the knowledge of $\tau_{max}$ does
not allow to make any sharp predictions on when the population will
reach convergence in the considered run.

Finally, Figure~\ref{f:relative_error} shows that the relative
standard deviation of all the relevant global quantities ($t_{max}$,
$t_{diff}$, $t_{conv}$ and $N_w^{max}$) decreases slowly as the system
size $N$ grows. In general, if the ratio $\sigma(x)/\la x \ra$ goes to
zero as $N$ increases, the system is said to exhibit self-averaging,
and this seems to be the case for the Naming Game. However, it is
difficult to draw a definitive conclusion, due to the large amount of
time needed to perform a significant number of simulation runs for
large values of $N$. Seemingly, the system seems to show
self-averaging for what concerns the peak height and time, but this
does not seem the case for the time of convergence. In any case, it is
worth mentioning that Lu, Korniss and Szymanski~\cite{lu2006} conclude
that a slightly modified version of the Naming Game model does not
display self-averaging when the population is embedded in random
geometric networks.

\section{Convergence Word}\label{s:conv_word}

\begin{figure}[t]
\begin{center}
\includegraphics*[width=10.5cm]{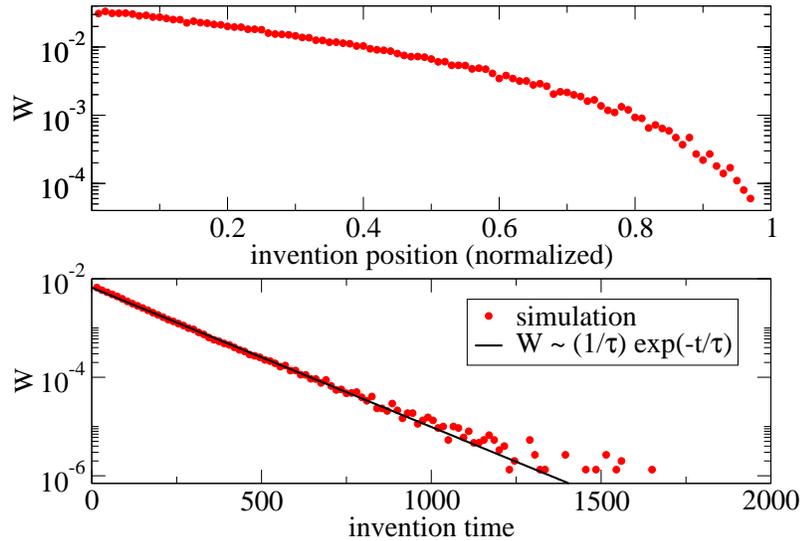}
\caption{{\small {\bf Word survival probability.} Top: The probability
    that a given word becomes the dominating one (i.e., the only one
    to survive when the system reaches the convergence state) is
    plotted as a function of its normalized invention position (see text
    for details). Early invention is clearly an advantageous
    factor. Bottom: the survival probability is now plotted in
    function of the invention time of words. The experimental
    distribution can be fitted by an exponential of the form $W \sim
    \; (1/\tau) \exp (-t/\tau)$, with $\tau \approx 150$. In both
    graphs, data have been obtained by $10^5$ simulation runs of a
    population made of $N=10^3$ agents.  }}
\label{f:winword_static}
\end{center}
\end{figure} 

\begin{figure}[t]
\begin{center}
\includegraphics*[width=10.5cm]{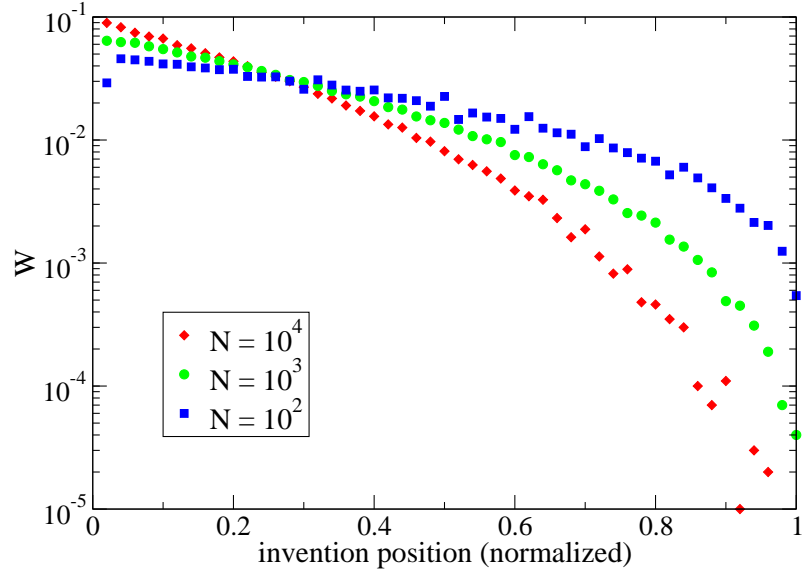}
\caption{{\small {\bf Role of the system size on the distribution of the
winning word.} The advantage of early invention increases in larger
populations.}} 
\label{f:winword_static_N}
\end{center}
\end{figure} 

As we have seen, the negotiation process leading agents to convergence
can be seen as a competition process among different words. Only one
of them will survive in the final state of the system. It is therefore
interesting to ask whether it is possible to predict, at some extent,
which word is going to dominate.

According to the Naming Game dynamical rules, the only parameter that
makes single words distinguishable is their creation time. Thus, it
seems natural investigating whether the moment in which a word is
invented can affect its chances of surviving. It turns out that this
is indeed the case, as it is shown in
Figure~\ref{f:winword_static}. The upper graph plots the probability
for a word to become the dominating one as a function of its normalized
creation position. This means that each word is identified by its
creation order: the first invented word is labeled as $1$, the second
as $2$ and so on. To normalize the labels, they are then divided by
the last invented word.  From Figure it is clear that early invented
words have higher chances of survival. The supremacy can be better
quantified if we plot the winning probability of a word as a function of
its invention time, as it is done in the bottom graph of
Figure~\ref{f:winword_static}. We find that data from simulations are
well fitted by an exponential distribution of the form $W=(1/ \tau)
exp (-t/ \tau)$, indicating that the advantage of early invention is
indeed quite strong.

Finally, an interesting question concerns the behavior of the winning
probability distribution as a function of the system size $N$. In
Figure~\ref{f:winword_static_N} we show the distributions as a function
of normalized labels described above for three different system sizes,
$N=10^2$, $N=10^3$ and $N=10^4$. The advantage of earlier creation
increases with the system size, but our data do not allow clear
predictions about the behavior of the distribution in the
thermodynamic, $N \rightarrow \infty$, limit. We might speculate that the
distribution collapses into a Dirac's delta of the first invented
word~\cite{Szathmary_private}.

\section{Symmetry Breaking - A controlled case}\label{s:2words}

In the previous sections we have seen that the winning word is chosen
by a symmetry breaking process (section~\ref{sym_break1}). This is
true even if, as we have seen in section~\ref{s:conv_word}, early
invention increases the probability for a word to impose
itself. Indeed, if we start with an artificial configuration in which
each agent has a different word in its inventory, i.e., if we remove
the influences of the invention process, the process still ends up in
the usual agreement state (data not shown).

\begin{figure}[t]
\begin{center}
\includegraphics*[width=10.5cm]{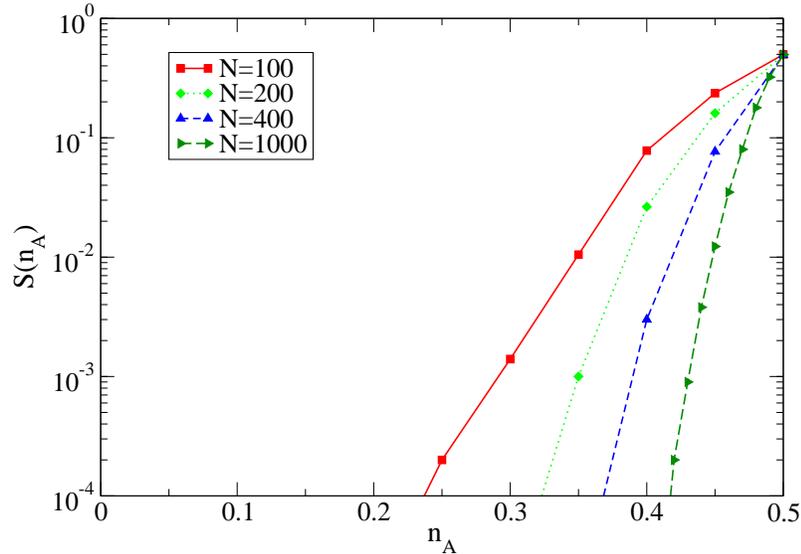}
\caption{{\small {\bf Resistance against invasion.} Two populations
    that converged separately on conventions $A$ and $B$ merge. In
    Figure it is plotted the probability $S(n_A)$ that convention $A$
    becomes the final accepted convention of the new population,
    versus the normalized size $n_A$ (where $n_A+n_B=1$) of the
    original population of $A$ spreaders.  As the total population
    size increases, the probability for the initially less diffused
    convention to impose itself decreases, as predicted by
    equations~(\ref{eq:2words}).}}
\label{f:sym_break}
\end{center}
\end{figure} 

In particular, we can concentrate on the case in which there are only
two words at the beginning of the process~\cite{baronchelli_ng_trans},
say $A$ and $B$, so that the population can be divided into three
classes: the fraction of agents with only $A$, $n_A$, the fraction of
those with only the word $B$, $n_B$, and finally the fraction of
agents with both words, $n_{AB}$ (see also~\cite{castello2006} for a
similar model). Describing the time evolution of the three species is
straightforward:

\bea \label{eq:2words}
{dn}_A/dt \, &=& \, - n_An_B + n_{AB}^2 + n_An_{AB} \\
\nonumber {dn}_B/dt \, &=& \, - n_An_B + n_{AB}^2 + n_Bn_{AB} \\
\nonumber {dn}_{AB}/dt \, &=& +2 n_An_B - 2n_{AB}^2 -
(n_A+n_B)n_{AB} \eea
 
The meaning of the different terms of the equations is clear. For
instance, for ${dn}_A/dt$ we have that $-n_An_B$ considers the
case in which an agent with the word $B$ transmits it to an agent with
the word $A$, $n_{AB}^2$ takes into account the probability that two
more agents with only the $A$ word are created if two agents with both
words happen to have a success with $A$, and $n_An_{AB}$ is due
to the probability that an agent with only $A$ has a success speaking
to an agent with both $A$ and $B$.
 
The system of differential equations~(\ref{eq:2words}) is
deterministic.  It has three fixed points in which the system can
collapse depending on initial conditions. If $n_A(t=0) > n_B(t=0)$
[$n_B(t=0) > n_A(t=0)$] then at the end of the evolution we will have
the stable fixed point $n_A=1$ [$n_B=1$] and, obviously,
$n_B=n_{AB}=0$ [$n_A=n_{AB}=0$]. If, on the other hand, we start from
$n_A(t=0) = n_B(t=0)$, then the equations lead to $n_A=n_B=2n_{AB}=b$,
with $b \simeq 0.18$.~\cite{baronchelli_ng_trans} The latter situation
is clearly unstable, since any external perturbation would make the
system fall into one of the two stable fixed points. Indeed, it is
never observed in simulations due to stochastic fluctuations that in
all cases determine a symmetry breaking forcing a single word to
prevail.

Equations~(\ref{eq:2words}), however, are not only a useful example to
clarify the nature of the symmetry breaking process. In fact, they
also describe the interaction among two different populations that
converged separately on two distinct conventions. In this perspective,
eq.~(\ref{eq:2words}) predicts that the population whose size is
larger will impose its conventions. In the absence of fluctuations,
this is true even if the difference is very small: $B$ will dominate
if $\;n_B(t=0) = 0.5 + \epsilon \;$ and $\; n_A(t=0) = 0.5 - \epsilon
\;$, for any $\; 0< \epsilon \leq 0.5\;$ (we consider
$n_{AB}(t=0)=0$).  Figure~\ref{f:sym_break} reports data from
simulations in which the probability of success of the convention of
the minority group $n_A$, $S(n_A)$, was monitored as a function of the
fraction $n_A$ (where $n_A+n_B=1$).  The absence of fluctuations is
partly recovered as the total number of agents grows, and in fact it
turns out that, for any given $n_A<0.5$, the probability of success
decreases as the system size is increased. Following
eq.~(\ref{eq:2words}), in the thermodynamic limit ($N \rightarrow
\infty$) this probability goes to zero.

\section{Discussion and Conclusions}\label{s:ng_discussion}

The Naming Game is the simplest model able to account for the
emergence of a shared set of conventions in a population of
agents. The main characteristics
are~\cite{baronchelli_ng_first,baronchelli_thesis}:

\begin{itemize}
\item The negotiation dynamics between individuals: the interaction
  rules are asymmetric and feedback is an essential ingredient to
  reach consensus;
\item The memory of the agents: individuals can accumulate words, and
  only after many interactions they have to decide on the final word
  chosen;
\item The absence of bounds to the inventory size: the number of words
  is neither fixed nor limited.
\end{itemize}

\noindent All these aspects derive from issues in Artifical
Intelligence, namely to understand how an open population of
physically embodied autonomous robots could self-organize
communication systems grounded in the world~\cite{steels03}. The model
is also relevant for all cases in which a distributed group of agents
have to tacitly negotiate decisions, as in opinion spreading or market
decisions~\cite{castellano07}. Nevertheless the ingredients listed
above are absent from most of the well known opinion-dynamics
models. In the Axelrod model~\cite{axelrod_model}, for instance, each
agent is endowed with a fixed-size vector of opinions, while in the
Sznajd model~\cite{sznajd} or the Voter
model~\cite{ligget,krapivsky_voter1,krapivsky_voter2}, the opinion can
take only two discrete values, and an agent adopts deterministically
the opinion of one of its neighbors.  Deffuant et al.~\cite{deffuant}
model the opinion as a unique variable and the evolution of two
interacting agents is deterministic, while in the Hegselmann and
Krause model~\cite{hegselmann} opinions evolve as an averaging
process. Most of these models include in some way the concept of
bounded confidence, according to which two individuals do not interact
if their opinions are not close enough, something which is entirely
absent in the Naming Game.  Interestingly, a recently proposed
generalized version of the Naming Game, in which a simple parameter
rules the consolidation behavior of the agents after a game, shows a
non-equilibrium phase transition in which the final state can be
consensus (as in the model we have analyzed in this paper),
polarization (a finite number of conventions survives asymptotically)
or fragmentation (the final number of conventions scales with the
system size)~\cite{baronchelli_ng_trans}, thus showing some phenomena
also found for most opinion dynamics models.

Compared to earlier Semiotic Dynamics models of the Naming
Game~\cite{steels99collectiveLearning}, this paper has made two
contributions. The effort towards the definition of simple interaction
rules has helped to bring out the essential features needed to achieve
a consensus state. Remarkably, we have shown that the weights
typically associated with word-meaning pairs in all earlier Naming
Game models are not crucial. The simplification does not impinge on
the ability of the model to be used on embodied agents i.e., it does
not introduce a global observer or other forms of global knowledge.

Next, because of the simplicity of the presented model, we have been
able to perform a comprehensive analysis of its behavior which has
never been done with earlier models due to their complexity. We have
investigated the basic features of the process leading the population
to converge, and how the crucial quantities scale with system size. In
this context, we have also revealed a hidden timescale that rules
the transition between the initial state, in which there is no
communication among agents, and the final one, in which there is
global agreement. Then we have analyzed several other aspects of the
whole process, such as its properties of convergence, the relation
between single runs and averaged curves, and the different
probabilities for single words to impose themselves. We have also
studied the elementary case in which only two words are present in the
system, which can be interpreted as the merging of two converged
populations, that clarifies the role of stochastic fluctuations in the
convergence process. Although many of these results have been seen in
numerical simulations, we have here been able to perform for the first
time a mathematical analysis. In future work, the techniques we have
used will be applied to more complex forms of communication including
grammatical language for which some Artificial Intelligence models
already exist~\cite{steels06grammarParsing}.

\section{Acknowledgments}
The authors wish to thank A. Barrat, E. Caglioti, C. Cattuto, L. Dall'Asta,
M. Degli Esposti, M. Felici, G. Gosti, A. Puglisi and V.D.P. Servedio
for helpful and interesting discussions.  A. Baronchelli acknowledges
support from the DURSI, Generalitat de Catalunya (Spain) and from
Spanish MEC (FEDER) through project No: FIS 2007-66485-CO2-01.  This
work was partially supported by the EU under contract IST-1940
(ECAgents) and IST-34721 (TAGora). The ECAgents and TAGora projects
are funded by the Future and Emerging Technologies program (IST-FET)
of the European Commission.

\end{document}